\documentstyle[version2,preprint,aps]{revtex}
\begin{document}
\draft
\begin{title}
Collective excitations in double-layer quantum-Hall systems
\end{title}
\author{A. H. MacDonald}
\begin{instit}
Department of Physics, Indiana University, Bloomington, Indiana 47405
\end{instit}
\author{Shou-Cheng Zhang}
\begin{instit}
IBM Research Division, Almaden Research Center, San Jose, CA 95120
\end{instit}

\begin{abstract}

We study the collective excitation spectra of double-layer
quantum-Hall systems using the single mode approximation. The
double-layer in-phase density excitations are similar to those of a
single-layer system. For out-of-phase density excitations, however,
both inter-Landau-level and intra-Landau-level double-layer modes have
finite dipole oscillator strengths. The oscillator strengths at long
wavelengths for the latter transitions are shifted upward by
interactions by identical amounts  proportional to the interlayer
Coulomb coupling.  The intra-Landau-level out-of-phase mode has a gap
when the ground state is incompressible except in the presence of
spontaneous inter-layer coherence.  We compare our results with
predictions based on the Chern-Simons-Landau-Ginzburg theory for
double-layer quantum Hall systems.

\end{abstract}


\narrowtext
\newpage

\section{Introduction}

Recent progress in the epitaxial growth of layered semiconducting
material has made it possible to fabricate samples containing two
high-mobility two-dimensional electron layers (2DEL) with a layer
separation, $d$, comparable to the typical separation of electrons
within a layer.  The experimental observation of a quantum Hall
effect\cite{exp} in these systems at total Landau level filling factor
equal to one-half has confirmed long-standing theoretical predictions
of novel quantum Hall effects due to inter-layer coupling\cite{halp}.
The quantum Hall effect occurs whenever the ground state of the system
is incompressible implying that there is a gap for charged
excitations. In this article we present a  discussion, based on the
single-mode approximation (SMA), of both intra-Landau-level and
inter-Landau-level (cyclotron) collective (neutral) excitations of the
incompressible ground  states.

We restrict our attention to the case where the two 2DEL's are
identical and tunneling between them may be neglected.  In this case
the number of electrons in each layer is a good quantum number and
collective modes corresponding to in-phase and out-of-phase density
oscillations in the two layers decouple. To use the SMA at strong
fields we evaluate separately projected  oscillator strengths for
intra-Landau-level and inter-Landau-level transitions for both in-phase
and out-of-phase density oscillations.  We work in the strong magnetic
field limit where the characteristic interaction energy  is much
smaller than the Landau level energy separation;
$\hbar \omega_c = \hbar e B / m^{*} c$ where $m^*$ is the effective
mass of the two-dimensional electron system.  We derive expressions
which relate these oscillator strengths to the intra-layer and
inter-layer correlation functions in the ground state. The SMA assumes
that each of the four oscillator strengths results from the excitation
of a single corresponding collective mode. In this case the
interaction contribution to the excitation energy is the quotient of
the interaction contribution to the projected oscillator strength and
a corresponding projected  static structure factor.  The SMA has
previously been applied successfully to study collective excitations
of an incompressible state for single-layers\cite{gmp,mog,caveat1}
systems.
In that case the intra-Landau-level excitation has oscillator strength
and structure factor both proportional to $k^4$ in the long wavelength
limit\cite{gmp} while the inter-Landau level excitation has oscillator
strength proportional to $k^2$ and is not shifted by electron-electron
interactions at long wavelengths. (In the long-wavelength (dipole)
approximation only the part of the projected oscillator strength which
goes like $k^2$ is included and we will refer to the portion of  the
oscillator strength which goes like $k^2$ as $k \to 0$ as the  dipole
contribution.) These behaviors result from the conservation of
particle number and invariance under translation; a long wavelength
electromagnetic field couples only to the cyclotron motion of the
center of mass of the system\cite{kohn}. We find similar behavior here
for the in-phase density excitations of the double-layer system.  For
the out-of-phase density excitations in the double-layer systems,
however, both intra-Landau-level and inter-Landau-level excitations
have dipole contributions to their oscillator strengths.  The
inter-Landau-level excitation is shifted from $\hbar \omega_c$ and the
intra-Landau-level excitation energy is generically finite at long
wavelengths.  An important exception occurs when coherence develops
spontaneously between the two-layers (see below). This difference
between the behavior of in-phase and out-of-phase modes is due to the
absence of invariance under relative translations of the two-layers.
Spontaneous interlayer coherence changes  the excitation spectrum
qualitatively because it requires fluctuations in the particle-number
difference between the two layers.

Recently, the Chern-Simons-Landau-Ginzburg (CSLG) theory\cite{cslg1}
of the single-layer quantum Hall effect has been generalized to the
double-layer case\cite{renn,cslgwz,cslge}. The CSLG theory starts from
the singular gauge transformation\cite{sgt} in which particles in one
layer see $m$ flux-quanta attached to  particles in the same layer and
$n$ flux-quanta attached to particles to the other layer.  Here $m$
and $n$ are integers and $m$ is odd so that  the transformation
changes the electrons into Bose particles.   At zeroth order the CSLG
theory replaces the resulting flux-density by its spatial average;
fluctuations are treated perturbatively. In the CSLG theory the
incompressible states are associated with situations where the
resulting Bose particles experience zero flux-density on average; {\it
i.e.} the statistical flux from the singular-gauge transformation
($m+n$ quanta per electron) must exactly cancel the flux from the
physical magnetic field.  This condition is satisfied when there are
$m+n$ quanta of physical magnetic field per electron, {\it i.e.} when
the double-layer  system has total Landau level filling factor
($\nu_T$) equal to $2/(m+n)$.  Incompressible states
occur for double-layer systems at a particular filling factor only if
the layer separation is in the appropriate range. It can be argued
that interlayer interactions of the appropriate  strength suppress
fluctuations in the statistical magnetic field and validate the CSLG
theory approach.  It is interesting that in the CSLG theory
random-phase-approximation\cite{fredkin} the many-body ground state
wavefunction is identical to Halperin's
$(m,m,n)$ wavefunction which is expected on the basis of direct
microscopic considerations to be approximately correct  at
$\nu_T=2/(m+n)$ when the layer separation is in the appropriate  range
and the ground state is incompressible.

In the CSLG theory random-phase-approximation, the long wavelength
out-of-phase density excitation mode has energy $\hbar\omega_c
(m-n)/(m+n)$. A gapless mode is thus obtained for the case where
$m=n$.  It has been claimed that this gapless mode implies
superfluidity in the double-layer system\cite{cslgwz,cslge}.  The
gapless mode in the CSLG theory exhausts the dipole oscillator
strength, {\it i.e.} the part of the oscillator strength proportional
to $k^2$ at long-wavelength.   However, the results derived below
prove that part of the dipole oscillator strength is associated with
inter-Landau-level excitations which occur at energies near $\hbar
\omega_c$, even when $m=n$.  This serious deficiency of the
random-phase-approximation in the CSLG theory can be resolved only by
considering the mixing of Gaussian fluctuations and vortex
excitations\cite{inprogress}. Moreover, we argue below that although
gapless out-of-phase modes associated with a spontaneously broken
continuous symmetry can occur in double-layer systems they do not
imply superfluid behavior.

Our article is organized as follows.  In Section II we  present our
SMA expressions for the inter-Landau-level cyclotron oscillator
strengths in the double-layer systems.  In Section III we present the
corresponding expressions for the lower energy intra-Landau-level
oscillator strengths.  In Section IV we present results for  the
excitation energies obtained using the SMA expressions and
approximate ground state correlation functions of the  double-layer
systems.  Here we discuss the import of our microscopic results for
the Chern-Simons-Landau-Ginzburg theory of double-layer systems and
emphasize special features of the correlation  functions which hold
only in the case where spontaneous interlayer coherence occurs in the
double-layer system. We conclude in Section V with a brief summary of
our results.

\section{Oscillator Strengths for Cyclotron Modes}

In the following two sections we assume that the magnetic field is
strong enough that the Landau level filling factor per layer  is less
than two, that the ground state has been completely  spin-polarized by
that magnetic field, and that Landau level mixing by interactions can
be neglected.  We use the  approach and employ the notation of
reference\cite{mog}.  (In particular we use $\ell \equiv (\hbar c /e
B)^{1/2}$ as the unit of length and  use complex number notation for
two-dimensional vectors.)  In that paper the Fourier components of the
density operator  for a single-layer are separated into contributions
coming from transitions between different pairs of Landau levels:
\begin{equation}
\rho_{k} = \sum_{n',n} \rho_{k}^{n',n}
\label{deceq:1}
\end{equation}
where
\begin{equation}
\rho_k^{n',n} = \sum_i |n'\rangle_i\langle n|_i G^{n',n}(k) B_i(k),
\label{deceq:2}
\end{equation}
the $G^{n',n}(k)$ are related to Laguerre polynomials\cite{mog}, the
sum over $i$ is over particle labels, and $B_i(k)$ comes from the
projection of $\rho_k$ onto a single Landau level and operates on the
intra-Landau-level degree of freedom of  particle $i$.  The
commutators which appear below are evaluated by  using
Eq.~[\ref{deceq:2}] and the identity,
\begin{equation}
B_i(k_1) B_i(k_2) = \exp (k_1^{*} k_2/2) B_i(k_1+k_2)
\label{deceq:3}
\end{equation}

The Hamiltonian of the double-layer system in the absence of
interlayer tunneling may be written in the following form which is
convenient for subsequent calculations:
\begin{eqnarray}
H & = & \hbar\omega_c \sum_i [a_i^{\dagger}(L) a_i(L) +
a_i^{\dagger}(R) a_i(R)] + \nonumber \\
  && {\textstyle{ 1 \over 2}}  \sum_q [V_q^{LL} \rho_q(L)\rho_{-q}(L)+
V_q^{RR} \rho_q(R)\rho_{-q}(R)+ 2  V_q^{LR} \rho_q(L)\rho_{-q}(R)]
\label{eq:1}
\end{eqnarray}
where $a_i(L)$ is the Landau level lowering operator for particle $i$
in the left (L) layer, $V_q^{LL} = V_q^{RR}$ is the intra-layer
Coulomb interaction, $V_q^{LR}$ is the inter-layer Coulomb
interaction, and $\rho_q(X)$ is the density operator for layer X.
(For explicit calculation below we ignore the finite thickness of the
2D layers so that $V_q^{LL} = 2 \pi e^2 /q$ and $V_q^{LR}=\exp (-q d)
V_q^{LL}$.)  In Eq.~[\ref{eq:1}] the  Hamiltonian includes infinite
constant terms corresponding to the self-interaction of each electron
in the system.  It is  convenient to retain these terms so that the
interaction terms  in the Hamiltonian can be expressed in terms of
density operators.   Since, in all subsequent calculations, the
Hamiltonian enters only in commutators, these non-physical constant
terms never contribute. The operators which generate in-phase and
out-of-phase inter-Landau level transitions are given by
$\Omega_k^{\pm} =  (\rho_k^{10}(L) \pm \rho_k^{10}(R) )/ \sqrt{2}$,
where $\rho_k^{10}(L)$ is the part of the density-operator associated
with transitions from the 0th to the 1st Landau levels.

The projected oscillator strengths for inter-Landau-level excitations
are defined by
\begin{equation}
f^{\pm}_{1,0}(k) \equiv \sum_n (E_n-E_0) |\langle n | \Omega_k^{\pm} |
0 \rangle |^2
\label{deceq:5}
\end{equation}
where $|n\rangle$ is an excited state of the system. Note that only
excitations in which the kinetic energy has  been raised by $\hbar
\omega_c$ can contribute. The oscillator strengths can be evaluated by
expressing them in terms of the commutator of the Hamiltonian with
$\Omega_k^{\pm}$.  $[T, \Omega_k^{\pm}] = \hbar\omega_c
\Omega_k^{\pm}$, where $T$ is the kinetic part of the Hamiltonian. The
evaluation of the commutator with the interaction part of the
Hamiltonian requires a lengthy calculation which leads to the
following result:
\begin{eqnarray}
\Delta_k^{\pm} & \equiv & \langle 0|(\Omega_k^{\pm})^{\dagger} [H,
\Omega_k^{\pm}]|0 \rangle \ / \ \langle 0|(\Omega_k^{\pm})^{\dagger}
\Omega_k^{\pm} |0 \rangle  = f^{\pm}_{1,0}(k) / s^{\pm}_{1,0} (k)
\nonumber \\
 & = &  \hbar \omega_c + \bar\rho |G^{10}(k)|^2 e^{-|k|^2/2} (V_k^{LL}
\pm V_k^{LR})\nonumber \\
 & + & \int {d^2 \vec q \over (2 \pi)^2 } V_q^{LL} [ G^{11}(q)
e^{(q^*k-kq*)/2} h^{LL}(q) - h^{LL}(q) + |G^{10}(q)|^2 e^{|k|^2/2+\vec
k \cdot \vec q} h^{LL}(k+q) ]\nonumber \\
 & + &  \int {d^2 \vec q \over (2 \pi)^2} V_q^{LR} [ G^{11}(q)
e^{(q^*k-kq*)/2} h^{LR}(q) - h^{LR}(q)\pm |G^{10}(q)|^2
e^{|k|^2/2+\vec k\cdot \vec q} h^{LR}(k+q) ].
\label{eq:4}
\end{eqnarray}
($G^{11}(q) = 1 - |q|^2/2$ and $|G^{10}(q)|^2=|q|^2/2$ and
$G^{10}(k)=-i k /\sqrt{2}$.) In Eq.~(\ref{eq:4})
\begin{equation}
  s^{\pm}_{1,0}(k) \equiv \langle 0|(\Omega_k^{\pm})^{\dagger}
\Omega_k^{\pm} |0 \rangle  =  N |G^{10}(k)|^2 e^{-|k|^2/2}
\label{eq:3}
\end{equation}
The $s^{\pm}_{1,0}(k)$ are the projected static structure factors for
inter-Landau level transitions:
\begin{equation}
s^{\pm}_{1,0}(k) = \sum_n |\langle n | \Omega_k^{\pm} | 0 \rangle |^2.
\label{deceq:4}
\end{equation}

The pair correlation functions $h^{XX'}(q)$ appearing in
Eq.~(\ref{eq:4}) are related to the static structure functions by
$s^{LL}(q)\equiv \langle ~0|\rho_q(L)\rho_{-q}(L)|0\rangle = N [1 +
h^{LL}(q) + \bar\rho (2\pi)^2 \delta^2(\vec q)]$ and $s^{LR}(q)
\equiv \langle 0|\rho_q(L)\rho_{-q}(R)|0\rangle  = N [h^{LR}(q) +
\bar\rho (2\pi)^2 \delta^2(\vec q)]$ where $\bar\rho$ is the average
density in each layer.  $\Delta_k^{\pm}$ is a projected
oscillator-strength-weighted average  of the in-phase and out-of-phase
inter-Landau-level excitation energies.  At long wavelength all the
oscillator strength is typically contributed by a single collective
mode. If this is the case $\Delta_k^{\pm}$ gives the energy of the
mode; in the single mode approximation this is assumed to be true at
all wavevectors of interest.

We see from these equations that the energy of the in-phase density
mode approaches the cyclotron energy $\hbar\omega_c$ in the $k \to 0$
limit, in agreement with Kohn's theorem.  (Since interactions are
invariant under simultaneous translations in both layers the proof for
a single-layer system\cite{kohn} trivially generalizes to the case of
the two-layer in-phase mode.) However the energy of the out-of-phase
density excitation is shifted from the cyclotron energy in the long
wavelength limit:
\begin{equation}
\Delta^-(k=0) = \hbar\omega_c - \int {d^2 q \over (2\pi)^2}
 q^2 V_q^{LR} h^{LR}(q)
\label{eq:9}
\end{equation}
The shift always increases the mode frequency as we will prove in the
next section.  This result is expected since interactions are not
invariant under relative translations of the two-layers. Kohn's
theorem cannot be generalized to this case and the energy of the mode
is shifted by an amount proportional to the symmetry breaking
interaction.

\section{Oscillator Strengths for Magnetoroton Modes}

Next we investigate the oscillator strengths the for intra-Landau-level
excitations\cite{baps} of double-layer systems.  As in the
inter-Landau-level excitation case, it will frequently be the case
that the intra-Landau-level oscillator strength,
\begin{equation}
\bar f^{\pm}(k) \equiv \sum_n (E_n - E_0) |\langle n | \Lambda_k | 0
\rangle |^2
\label{deceq:6}
\end{equation}
is dominated by the contribution from a single collective mode. The
corresponding intra-Landau-level projected static structure factors
are defined by
\begin{equation}
\bar s^{\pm}(k) \equiv \sum_n |\langle n | \Lambda_k | 0 \rangle |^2
= \langle 0 | \Lambda_{-k} \Lambda_k | 0 \rangle.
\label{deceq:7}
\end{equation}
Following Ref.\cite{gmp} we call these intra-Landau-level collective
modes magnetorotons.  As in the previous section magnetoroton energies
can be estimated by comparing corresponding oscillator strengths and
static structure functions. In Eq.[\ref{deceq:6}] $\Lambda_k^{\pm}
\equiv  (\rho_k^{00}(L) \pm \rho_k^{00}(R) ) / \sqrt{2}$ where
$\rho_k^{00}(L)$ is the density operator projected to the lowest
Landau level.  $\bar f^{\pm}(k)$ is evaluated by  expressing it in
terms of a commutator.  In this case one finds, again after a lengthy
calculation, that
\begin{eqnarray}
\bar f_k^{\pm} & \equiv &  \langle 0|(\Lambda_k^{\pm})^{\dagger} [H,
\Lambda_k^{\pm}]|0\rangle\nonumber \\
 & = & {N \over 2} \int {d^2 \vec q \over (2 \pi)^2 }  V_q^{LL} [2
e^{-|k|^2/2}
[ \cos((\vec k \times \vec q)_z) -1] \bar s^{LL}(q) - (e^{q k^*/2} - e^{k
q^*/2})^2 \bar s^{LL}(k+q)]\nonumber \\
 & + & {N \over 2} \int {d^2 \vec q \over (2 \pi)^2 } V_q^{LR} [ 2
e^{-|k|^2/2}
[ \cos((\vec k \times\vec q)_z) -1] \bar s^{LR}(q) \mp (e^{q k^*/2} - e^{k
q^*/2})^2 \bar s^{LR}(k+q)].
\label{eq:11}
\end{eqnarray}
The projected structure factors appearing in Eq.~(\ref{eq:11}) are
related to pair correlation functions by $\bar s^{LL}(q) = h^{LL}(q)
+e^{-|q|^2/2}$ and $\bar s^{LR}(q) = h^{LR}(q)$.  The structure
factors associated with the operators $\Lambda_k^{\pm}$ are
\begin{equation}
\bar s_k^{\pm}  \equiv
N^{-1} \langle 0|(\Lambda_k^{\pm})^{\dagger} \Lambda_k^{\pm}|0\rangle
  =   \bar s^{LL}(k) \pm \bar s^{LR}(k).
\label{deceq:8}
\end{equation}
The single-mode approximation collective mode energies for the two
intra-Landau-level modes are then given by $\bar f_k^{\pm}/ \bar
s_k^{\pm}$.

Expanding $\bar f_k^{\pm}$ for small $k$ we find that $\bar f_k^+ \sim
k^4$ whereas
\begin{equation}
 \bar f_k^- = - N (k^2/2)  \sum_q q^2 V_q^{LR} h^{LR}(q) + o(k^4)
\label{eq:12}
\end{equation}
It is interesting that for the out of phase mode the interaction
contributions to the dipole ($\propto |k|^2$) portions of the
intra-Landau-level and inter-Landau-level oscillator strengths are
identical. (For the inter-Landau-level case the dipole part of the
oscillator strength is $N |G^{10}(k)|^2 (\Delta^{-}(k=0)-\hbar
\omega_c)$. By definition $f_k^-$ is positive definite so that
$\Delta^{-}(k=0) - \hbar \omega_c$ must also be positive.  In the
single-mode approximation the out-of-phase cyclotron mode is always
shifted to higher energy by electron-electron interactions.  The
situation is similar to that  for the effect of disorder on the
vibration modes of the Wigner  crystal at strong magnetic fields where
pinning the  crystal shifts both intra-Landau-level and
inter-Landau-level modes upward by the same amount\cite{macdunpub}.
We remark that the unprojected oscillator strengths,
\begin{equation}
f_k^{\pm} = \sum_n |\langle n |(\rho_q(L) \pm \rho_q(R))/\sqrt{2} | 0
\rangle |^2 (E_n-E_0) = N{ \hbar^2 k^2 \over 2 m^* }
\label{eq:jan18}
\end{equation}
are not shifted by electron-electron interactions.  For the in-phase
mode the dipole contribution to the unprojected oscillator strength is
exhausted by the Kohn mode.  For the out-of-phase mode the
unprojected oscillator strength is independent of interactions and it
appears at first sight that the interaction corrections to the dipole
oscillator strengths for intra-Landau-level and inter-Landau-level
modes should cancel.  (Excitations with a Landau level index change
larger than one do not contribute to the dipole oscillator strength.)
However, the sum of the projected oscillator strengths does not equal
the unprojected oscillator strength to leading order in
electron-electron interactions; an additional contribution comes from
Landau level mixing which alters the projected structure for  the
inter-Landau level excitations by an amount proportional to
$(e^2/\ell) / (\hbar \omega_c)$.

{}For a single-layer\cite{gmp} system conservation of total particle
number and invariance under translations guarantee that  the projected
structure factor $\sim k^4$ for small $k$. As in the
inter-Landau-level case, the conservation laws which control the
long-wavelength behavior in a single layer have analogs for the
in-phase two layer excitations and, as discussed in the following
paragraph $\bar s^{+}_k \sim k^4$ at long wavelengths.  A dipole
oscillator strength appears only in the inter-Landau-level mode. The
relevant quantities for the out-of-phase mode are the particle-number
difference, which is conserved, and relative  translations of the two
layers, under which the Hamiltonian is {\it not} invariant in the
presence of interlayer interactions. {}From this we infer (see below)
that, unless the system exhibits  spontaneous interlayer
coherence, $\bar s_k^- \sim k^2$.   The dipole oscillator strength is
split between inter-Landau-level and intra-Landau-level modes.  From
these $k \to 0$ behaviors for $\bar s_k^{\pm}$ and $f_k^{\pm}$ we
conclude that both the in-phase and the out-of-phase intra-Landau-level
excitations of incompressible states within the lowest Landau level
will have a finite gap. An exception occurs when the two-layers have
spontaneous interlayer coherence as discussed in the next section.

In the following paragraphs we explicitly discuss relationships
between the  small wavevector limits of the projected static structure
factors and  conserved quantities in the double-layer system.  We
closely follow the corresponding discussion for the single-layer
system\cite{gmp}.  From their definitions it is easy to show that for
equal average density ($\bar \rho = \nu_T/ 4 \pi$) in the layers the
pair correlation functions  are related to their real space
counterparts by
\begin{equation}
h^{AB}(q)= { \nu_T \over 4 \pi}
\int d^2 \vec r \, h^{AB}(r) \exp (i \vec q \cdot \vec r).
\label{eq:jan2}
\end{equation}
where $h^{AB}(r)= (\bar \rho)^{-2} \, n^{AB}(0,\vec r)-1$.  Here
$n^{AB}(\vec r_1, \vec r_2)$ is a the two-point distribution function
with one particle in layer A and the other particle in layer B. For an
isotropic liquid, $n^{AB}(\vec r_1, \vec r_2)$  is dependent only on
the distance between the two points.  It will be convenient to use the
symmetric gauge in which the  single-particle states in the lowest
Landau level have the  form:
\begin{equation}
\phi_m(z) = {z^m \exp (- \bar z z/4) \over (2 \pi 2^m m!)^{1/2}}
\label{eq:jan1}
\end{equation}
with non-negative angular momentum $m$. This gauge is convenient
because at the  (arbitrarily chosen) origin only the $m=0$
wavefunction is non-zero. Expressing the two-point function in second
quantized and  using the isotropy of the fluid then gives
\begin{equation}
h^{LL}(r) = \nu^{-2} \sum_{m=0}^{\infty} \big[
\langle \hat n_{mL} \hat n_{0L} \rangle  (1 - \delta_{m,0})
 - \langle \hat n_{mL} \rangle  \langle \hat n_{0L}\rangle]
 {1 \over m!} (r^2/2)^m exp(-r^2/2)
\label{eq:jan2b}
\end{equation}
and
\begin{equation}
h^{LR}(r) = \nu^{-2} \sum_{m=0}^{\infty} \big[
\langle \hat n_{mL} \hat n_{0L} \rangle
 - \langle \hat n_{mL} \rangle  \langle \hat n_{0L}\rangle]
 {1 \over m!} (r^2/2)^m exp(-r^2/2).
\label{eq:jan19}
\end{equation}
In Eq.~(\ref{eq:jan2b}) and Eq.~(\ref{eq:jan19}) $\nu \equiv \nu_T/2$
is the filling factor per layer.  It follows from these equations that
\begin{equation}
{\nu \over 2 \pi} \int d^2 \vec r \, [h^{LL}(r) \pm h^{LR}(r)]
 = -1+ \langle \langle (\hat N_L \pm \hat N_R) \hat n_{0L} \rangle
\rangle
\label{eq:jan20}
\end{equation}
and
\begin{equation}
{\nu \over 2 \pi} \int d^2 \vec r \, (r^2/2) [h^{LL}(r) \pm h^{LR} (r)]
 = -1+  \langle \langle (\hat M_L + \hat N_L \pm \hat M_R \pm \hat M_L
) n_{0L} \rangle \rangle.
\label{eq:jan21}
\end{equation}
Here $\langle \langle A B \rangle \rangle \equiv \langle 0 | A B | 0
\rangle - \langle 0 | A | 0 \rangle \langle 0 | B | 0 \rangle$ for
operators $A$ and $B$, $\hat N_X = \sum_m \hat n_{mX}$ is the total
number operator for layer $X$, $\hat M_X = \sum_m m \hat n_{mX}$ is
the total angular momentum operator for layer $X$ and $\hat n_{mX}$ is
the number operator for the single-particle state with angular momentum
$m$ in layer $X$.

The second term on the right hand side of  Eq.~(\ref{eq:jan20})
vanishes if $N_L \pm N_R$ is a good quantum number while the second
term on the right hand side of  Eq.~(\ref{eq:jan21}) vanishes if both
$N_L \pm N_R$ and $M_L \pm M_R$ are good quantum numbers.  Since $N_L
\pm N_R$ commutes with the Hamiltonian the right hand side of
Eq.~(\ref{eq:jan20}) should equal $-1$, unless the ground state has a
broken symmetry.  As we  comment further below a broken symmetry
characterized by spontaneous phase coherence between the layers does
occur in double-layer  systems under appropriate circumstances and
when it occurs  it is accompanied by fluctuations in $N_L - N_R$.  The
Hamiltonian commutes with $M_L + M_R$ since it is invariant under
rotations of the all coordinates but does not commute with $M_L - M_R$
since it is not invariant under rotation of one layer with respect to
the other. It follows the right hand side of Eq.~(\ref{eq:jan21}) is
zero only for the case of the plus sign.  Expanding the plane wave
factor in  Eq.~(\ref{eq:jan2}) and using these sum rules  we may
conclude that the in-phase intra-Landau-level projected  structure
factor $\bar s^{+}_k \sim k^4$ always, whereas the out-of-phase
intra-Landau-level projected  structure factor $\bar s^{-}_k \sim k^2$
as long as  spontaneous interlayer phase coherence does not occur. We
emphasize that these results follow from general sum rules and are not
based on any particular approximate many-body wavefunction for the
incompressible ground state.   Since $\bar f^{+}_k \sim k^4$ and $\bar
f^{-}_k \sim k^2$ independent of the long wavelength  behavior of
$\bar s^{\pm}_k$ it follows that the in-phase intra-Landau-level
collective mode has a gap and the out-of-phase  intra-Landau-level
collective mode has a gap except in the  case of spontaneous
interlayer coherence.  This conclusion  rests on the assumption that
the generalized single-mode-approximation is valid at long wavelengths.

\section{Numerical Results and Discussion}

The above expressions allow us to estimate the collective mode
energies of a double-layer system given an approximation for the
ground state spatial correlations.  In Fig.~(\ref{fig:1}) we show
results obtained for a double-layer system with a total Landau level
filling factor $\nu_T = 1/2$ and a layer separation $d = 1.5 \ell$,
close to the effective layer separation value for which novel
double-layer fractional Hall effects have recently been
observed\cite{exp}.  Numerical calculations\cite{gsnum} have
established that the ground state at this value of
$d/l$ is accurately by the $(m,m,n)=(3,3,1)$ Halperin\cite{halp}
wavefunction and we have used the correlation functions\cite{rm,baps}
of that wavefunction to evaluate the oscillator strengths and
structure factors. For $k \to 0$ the in-phase inter-Landau-level mode
(the Kohn mode) is unshifted by interactions while the out-of-phase
mode is shifted to higher energies as discussed above. The shift,
which is a direct measure of inter-layer correlations,
should\cite{caveat2} be observable in cyclotron resonance experiments
in double-layer systems.  Note also that both in-phase and
out-of-phase intra-Landau level modes have a finite gap as expected
from the above discussion.

Similar results are shown in Fig.~(\ref{fig:2}) for a double-layer
system with $\nu_T=1$ and $d = \ell$ with  correlation functions
approximated by those of the
$(m,m,n)=(1,1,1)$ Halperin state.   This wavefunction is actually {\it
not} a good approximation to the ground state of a double-layer system
except for the limit $d \to 0$ as we discuss below and the estimates
obtained, at least  in the case of the intra-Landau-level modes, are
unreliable. A similar situation arises for $\nu_T=1/m$ where the
ground state as $d \to 0$ approaches the $(m,m,m)$ Halperin
wavefunction.  The
$(m,m,n)=(1,1,1)$ Halperin wavefunction is a single Slater determinant
and it is easy  to evaluate its correlation functions analytically:
$h^{LL}(k)=h^{LR}(k) = - \exp (-|k|^2/2)/2$; $\bar s^{+}(k) \equiv 0$;
$\bar s^{-}(k) = \exp  (- |k|^2/2 )$.   Because $\bar s^{+}(k)$
vanishes identically for this wavefunction, we cannot use it to
estimate the energy of the intra-Landau-level in-phase mode; the fact
that $\bar s^{+}(k)$ vanishes can be understood by noting that
$(1,1,1)$ is the only state with equal layer population at $\nu_T=1$
which is a total isospin\cite{gapless,halp} eigenstate with eigenvalue
$N/2$.  (States in a definite layer  are taken to be eigenstates of
the $\hat z$ component of an isospin operator.)  The actual ground
state at $d/\ell \ne 0$ is {\it not} a total isospin eigenstate because
of the layer-dependence of the electron-electron interactions but is
still expected\cite{cslgwz,cslge,gapless,myzgmzy,harfok} to show a
spontaneous symmetry breaking corresponding to the development of
inter-layer coherence in the absence of tunneling, or equivalently to
isospin polarization in the
$\hat x- \hat y$ plane.  This suggests that $(1,1,1)$ is not a good
approximation to the ground state at any $d/\ell \ne 0$; in fact, we
expect\cite{myzgmzy} that for the true ground state at $\nu_T = 1$,
$\bar s^{-}(k) \sim k$ so that the long-wavelength dispersion relation
of the out-of-phase intra-Landau level mode is linear\cite{gapless}
rather the quadratic result obtained with the above correlation
functions.  The inter-Landau-level modes are not very sensitive to the
long wavelength behavior of the correlation functions and are probably
still accurately estimated using the $(1,1,1)$ state correlation
functions.

We are now able to compare our results for the collective modes with
the CSLG theory of the double-layer system\cite{cslgwz,cslge}. In the
CSLG random-phase-approximation, the in-phase and out-of-phase density
correlation functions are given by:
\begin{equation}
\rho_{+} (\omega, q) = \frac {\bar\rho q^2 /m^*} {\omega^2 -
\omega_{+}^2}  \ \ \ , \ \ \ \rho_{-} (\omega, q) = \frac {\bar\rho
q^2 /m^*} {\omega^2 - \omega_{-}^2}
\label{eq:14}
\end{equation}
where $\omega_{+} = \omega_c = eB/m^* c$, $\omega_{-} = \omega_c
(m-n)/(m+n)$ and $m^*$ is the effective band mass of the electrons.
The in-phase density mode can be clearly identified with the Kohn
mode\cite{lz}; and its energy is not shifted by higher order
corrections in the CSLG theory. However there are difficulties in
identifying the out-of-phase mode in this theory.  From
Eq.~(\ref{eq:12}) we see that the intra-Landau level mode should have
a dipole oscillator strength proportional to $V^{LR} h^{LR}$.  One
might therefore be tempted to identify $\omega_{-}$ with the
out-of-phase intra-Landau-level mode. Then for the case of the $m=n$
random-phase-approximation, $\omega_{-} = 0$, and it would then be
tempting to identify this mode with the gapless intra-Landau level
mode obtained earlier in the SMA analysis when the ground state is
close to the $m=n$ Halperin state. Unfortunately, the single
out-of-phase mode calculated within the double-layer CSLG theory
saturates the full dipole oscillator strength $\bar\rho q^2/m^*$. This
is not acceptable since an excitation within the lowest Landau level
can not contain explicit dependence on the band mass $m^*$.  The
second possibility is to  interpret the out-of-phase mode obtained in
equation (15) as an inter-Landau level mode.  In this case one is
faced with the difficulty that the mode energy is shifted downwards
from the cyclotron energy by an amount proportional to $\omega_c$,
whereas the SMA calculations show that it should be shifted upwards by
an amount proportional to the interlayer Coulomb energy.  In a future
publication\cite{inprogress} it will be shown that it is crucial to
include the mixing of the vortex excitations with Gaussian
fluctuations in the CSLG theory in order to resolve these difficulties.

It is interesting to observe\cite{bonesteel} that when the ground
state is given by the $(m,m,n)$ Halperin wavefunction with $m > n$,
the  oscillator strength weighted average excitation energy at long
wavelengths is $\omega_{-}$.  In the strong magnetic field limit where
$(m,m,n)$ can be a good approximation to the  ground state at
$\nu_T=2/(m+n)$ the oscillator strength weighted average excitation
energy is $\hbar \omega_c$ times the fraction of the total structure
factor which is  contributed by inter-Landau-level transitions.  For
the case  of the $(m,m,n)$ wavefunction it is easy to
show\cite{forrester} that that fraction is $(m-n)/(m+n)$.  It is our
belief that  the CSLG theory random-phase-approximation finds a
collective  out-of-phase mode at $\omega_{-}$ because it gives the
correct total out-of-phase oscillator strength and gives rise to the
$(m,m,n)$ ground state wavefunction  but incorrectly places all the
oscillator strength in a  single mode.   The oscillator strength
weighted average  excitation energy is reduced not because the
inter-layer out-of-phase mode is reduced in energy but rather because,
unlike the in-phase case, some of the dipole oscillator strength
resides in the low energy intra-Landau-level mode.

Finally we would like to remark on the question of Josephson effect in
the double-layer systems\cite{cslgwz,cslge}.  In the limit of vanishing
layer separation, the Hamiltonian is $SU(2)$ invariant in the isospin space.
However, when the layer separation is finite, the $SU(2)$ symmetry is
broken to an $U(1)$ symmetry, and the total isospin lies preferentially in the
$XY$ plane. The linearly dispersing
collective mode in the $m=n$ case is analogous to the magnon mode of a
two-dimensional XXZ ferromagnet\cite{gapless}. The spontaneous $U(1)$
symmetry breaking associated with the direction of the total isospin in the
$XY$ plane and the presence of the linearly dispersing collective mode
are similar to the properties of a superconductor and the isospin direction
can be associated with the phase order parameter of a superconductor.
This has led various
groups to the conjecture that a Josephson effect might exist in the
double-layer quantum Hall systems\cite{cslgwz,cslge}. However, we would
like to point out here that this phenomenon does not occur. While the
different isospin directions in the $XY$ plane are degenerate in the absence
of interlayer tunneling, this degeneracy is broken explicitly
(not spontaneously) by the tunneling, and
the total isospin of the system points preferentially in the $X$ direction.
In the presence of interlayer tunneling, there is a finite energy cost
to rotate the isospin in the $XY$ plane. Since the DC Josephson effect
requires a zero frequency mode in the phase dynamics of the superconducting
order parameter, the finite energy cost associated with the $XY$ isospin
dynamics shows that this effect is absent in the double-layer quantum Hall
system.

To be more explicit, let us consider the following Hamiltonian
which captures the essential physics we wish to discuss:
\begin{equation}
  H = -2 t \int d^2 r S_x(r)  + U \int d^2 r S_z^2 (r).
\label{josephson1}
\end{equation}
In Eq.~[\ref{josephson1}] the first term describes tunneling
and the second term
describes the capacitance energy in the double-layer systems.
For slowly varying $S_z (r)$ the capacitance energy
is the energy in the electric field between the two layers
created by charge transfer from one layer to another.  ($U \propto d$.)
The capacitance energy is the leading term in a spin-polarization
gradient expansion of the iso-spin dependent terms in the Hamiltonian.
The $S_i(r)$ for $i=x,y,z$ obey the usual spin commutation relations.
In this isospin language, the equation
of motion for the total isospin $S_i = \int d^2 r S_i(r)$ is given
by\cite{unpub}
\begin{equation}
  d S_y / dt = - 2 t S_z , \ \ d S_z / dt = 2 t S_y
\label{josephson2}
\end{equation}
While these equations are independent of the capacitance term, the first
equality in Ref.~[\ref{josephson2}] would have contributions from higher
order terms in the gradient expansion of the isospin-dependent terms in
the full double-layer Hamiltonian.
Since $S_z$ is nothing but the difference in the number of electrons in
the upper and lower layers, $d S_z / dt$ describes the tunneling current
between the two layers. Solving these equations we find that the current
current correlation function has a finite frequency pole at $\omega =
\pm 2 t$, and that
there is no zero
frequency delta function contribution to the current current
correlation which would be associated with the putative Josephson
effect\cite{myzgmzy,unpub}. We also notice that the capacitance energy
term in Eq.~[\ref{josephson1}]
does not affect the pole in the total current current correlation function.
This is a general feature also present in the case of a charged
superconductor, where the long ranged Coulomb energy gives rise to a
finite plasmon energy in the wave vector
$q\rightarrow 0$ limit but does not
contribute to the total current current correlation function at $q=0$.

\section{Summary}

In conclusion we have calculated explicitly the collective mode spectrum
for double-layer qauntum Hall systems within the single mode
approximation. The in-phase density mode shares the essential
properties of density modes in single-layer quantum Hall systems.
The cyclotron mode is unshifted by interactions in the long wavelength
limit and there is no dipole oscillator strength for intra-Landau-level
modes.  For the
out-of-phase mode, however,
the energy for inter-Landau-level transitions is
shifted upwards from the cyclotron energy at long wavelengths
by an amount determined by the
interlayer interaction strength. The intra-Landau-level transition
for the out-of-phase mode has an oscillator strength proportional to
$q^2$, and its energy is generically finite.  The exception occurs
when there is spontaneous symmetry breaking which establishes phase
coherence between the two-layers in the absence of tunneling.
In an isospin language this phase coherence corresponds to
easy-plane ferromagnetism and when it is present the out-of-phase
mode is gapless and has linear dispersion at long wavelengths.
We also compared the SMA calculation with the
predictions of the Chern-Simons-Landau-Ginzburg theory. While the CSLG
theory correctly captures the ground state properties and the excitation
spectrum for the in-phase mode, within the Gaussian approximation,
it incorrectly places all the dipole
oscillator strength for the out-of-phase excitation in a single mode
and the resulting energy spectrum differs from the predictions
of this SMA calculation. To cure
this short-coming in the CSLG theoey,
it is necessary to go beyond the simple Gaussian or
RPA approximation and include the interplay between the Gaussian and vortex
degrees of freedom.  Finally we have shown that the putative
Josephson effect in double-layer quantum Hall systems does not occur.

After this work was complete we learned of related work\cite{renn2} by
Renn and Roberts where some of our results for intra-Landau-level
collective modes are independently derived.

This work was supported in part by the National Science Foundation
under grant DMR-9113911.  The authors acknowledge stimulating and
instructing conversations with D.P. Arovas, S.M. Girvin, D.-H. Lee,
X.G. Wen, and Y.S Wu at the Aspen Center for  Physics and thank S.M.
Girvin for a careful reading of the compuscript.

\newpage

\figure{
Collective mode dispersion for a double-layer system at $\nu_T=1/2$
and $d/\ell=1.5$.  The energies of the inter-Landau-level modes are
measured from $\hbar \omega_c$.  The ground state is approximated by
the $(3,3,1)$ Halperin state.  The plotting symbols refer to the
following modes: plus (inter-Landau-level in-phase); cross
(inter-Landau-level out-of-phase); square (intra-Landau-level
in-phase); diamond (intra-Landau-level out-of-phase).
\label{fig:1}}

\figure{
Collective mode dispersion for a double-layer system at $\nu_T=1$ and
$d/\ell=1.0$ as calculated from the correlation functions of the
$(1,1,1)$ Halperin state. The plotting symbols have the same
definitions as in Fig.~(\ref{fig:1}).  The energies of the
inter-Landau-level modes are measured from $\hbar \omega_c$. The
intra-Landau-level mode energies are sensitive to the long wavelength
behavior of $\bar s^{\pm}(k)$ which is qualitatively in error at
$d/\ell \ne 0$ when the ground state is approximated by the $(1,1,1)$
Halperin state.
\label{fig:2}}


\begin{references}

\bibitem{exp} Y. W. Suen {\it et al.}, Phys. Rev. Lett. {\bf 68}, 1379
(1992); J. P. Eisenstein {\it et al.}, Phys. Rev. Lett. {\bf 68}, 1383
(1992).

\bibitem{halp} B. I. Halperin, Helv. Phys. Acta {\bf 56}, 75 (1983);
For a brief review of the fractional quantum Hall effect in
multi-layer two-dimensional electron systems see A.H. MacDonald,
Surface Science {\bf 229}, 1 (1990).

\bibitem{gmp} S. M. Girvin, A. H. MacDonald and P. M. Platzman, Phys.
Rev. B {\bf 33}, 2481 (1986).

\bibitem{mog} A. H. MacDonald, H. C. Oji and S. M. Girvin, Phys. Rev.
Lett. {\bf 55},  2208 (1985).

\bibitem{caveat1} The SMA is not reliable for compressible states or
for hierarchical incompressible states.

\bibitem{kohn} W. Kohn, Phys. Rev. {\bf 123}, 1242 (1961); C. Kallin
and B.I. Halperin, Phys. Rev. B {\bf 31}, 3635 (1985).

\bibitem{cslg1} For a review, see Shou-Cheng Zhang, Int. J. of Modern
Physics, B {\bf 6}, 25 (1992).

\bibitem{renn} S.R. Renn, Phys. Rev. Lett. {\bf 68}, 658 (1992).

\bibitem{cslgwz} X.G. Wen and A. Zee, Phys. Rev. Lett. {\bf 69}, 1811
(1992); X.G. Wen and A. Zee, Phys. Rev. B {\bf 47}, 2265 (1993).

\bibitem{cslge} Z.F. Ezawa  and A. Iwazaki, Int. J. of Mod. Phys. B,
{\bf 19}, 3205 (1992); Z.F. Ezawa and A. Iwazaki, Phys. Rev. B {\bf
47}, 7295 (1993); Z.F. Ezawa, A. Iwazaki, Phys. Rev. B {\bf 48}, 15189
(1993).

\bibitem{sgt} F. Wilczek, Phys. Rev. Lett. {\bf 48}, 1144 (1982).

\bibitem{fredkin} A. Lopez and E. Fradkin, Phys. Rev. Lett. {\bf 69},
2126 (1992); A. Lopez and E. Fradkin, Phys. Rev. B {\bf 47}, 7080
(1993).

\bibitem{inprogress}  A modified CSGL theory which accommodates the
microscopic results derived here has recently been developed. D.P.
Arovas, Dung-Hai Lee and Shou-Cheng Zhang, in preparation (1993).

\bibitem{baps} A brief discussion of these modes has been given
previously.  A.H. MacDonald, Mark Rasolt and Francois Perrot, Bull.
Am. Phys. Soc. {\bf 33}, 748 (1988).

\bibitem{macdunpub} A.H. MacDonald, unpublished notes.

\bibitem{unpub} A.H. MacDonald, and Shou-Cheng Zhang, unpublished
(1993).

\bibitem{gsnum} Song He, S. Das Sarma and X.C. Xie, Phys. Rev. B {\bf
47}, 4394 (1993); D. Yoshioka, A.H. MacDonald, and S.M. Girvin, Phys.
Rev. B {\bf 39}, 1932 (1989); T. Chakraborty and P. Pietil\" ainen,
Phys. Rev. Lett., {\bf 59}, 2784 (1987); E.H. Rezayi and F.D.M.
Haldane, Bull. Am. Phys. Soc. {\bf 32}, 892 (1987).

\bibitem{rm} The correlation functions were obtained by using a HNC
approximation as described in earlier papers;  Mark Rasolt, F. Perrot,
and A.H. MacDonald, Phys. Rev. Lett. {\bf 55}, 433 (1985); Mark Rasolt
and A. H. MacDonald, Phys. Rev. B34, 5530 (1986).

\bibitem{caveat2} It's strength will be reduced by a factor of $\sim
(d/\cos(\theta) \lambda)^2$ where $\lambda$ is the wavelength of the
infrared light and $\theta$ is the angle between the propagation
direction and the normal to the 2D layers.

\bibitem{gapless} A.H. MacDonald, P.M. Platzman, and G.S. Boebinger,
Phys. Rev. Lett. {\bf 65}, 775 (1990); Luis Brey, Phys. Rev. Lett.
{\bf 65}, 903 (1990); H.A. Fertig, Phys. Rev. B {\bf 40}, 1087 (1989).

\bibitem{myzgmzy} Kun Yang, Kyungsun Moon, Lian Zheng,  A.H.
MacDonald, S.M. Girvin, Shou-Cheng Zhang, and Daijiro Yoshioka, Phys.
Rev. Lett., {\bf 72}, 732 (1994).

\bibitem{harfok} For detailed calculations within the Hartree-Fock
approximation see R. C\^ ote, L. Brey, and A.H. MacDonald, Phys. Rev.
B {\bf 46}, 10239 (1992).

\bibitem{lz} D. H. Lee and Shou-Cheng Zhang, Phys. Rev. Lett, 66, 1220
(1991).

\bibitem{bonesteel} Nick Bonesteel, private communications (1993).

\bibitem{forrester} P.J. Forrester, and B. Jancovici, J. Phys. (Paris)
Lett. {\bf 45}, L583 (1984).

\bibitem{note} In the presence of the long-ranged Coulomb interaction,
the long wave length limit $q \to 0$ is discontinuous. The density
density correlation function in the limit $q \to 0$ has a finite gap
at the plasma energy. However, for a superconductor, the $q=0$
current-current correlation function still has a delta function peak
at zero frequency, unaffected by the long ranged Coulomb interaction.

\bibitem{renn2} S.R. Renn and B.W. Roberts, Phys. Rev. B {\bf 48},
10926 (1993).

\end{references}
\end{document}